\documentclass[useAMS,usenatbib]{mnras}
\usepackage[latin1]{inputenc}
\usepackage[english]{babel}
\usepackage{graphicx}
\usepackage{amsmath}
\usepackage{amsfonts}
\title[$\Lambda CDM$ periodic cosmology]{$\Lambda CDM$ periodic cosmology}
\author[Stéphane Fay]{Stéphane Fay\footnote{steph.fay@gmail.com}\\
Palais de la Découverte\\
Astronomy Department\\
Avenue Franklin Roosevelt\\
75008 Paris\\
France}
\date{Accepted XXX. Received YYY; in original form ZZZ}
\pubyear{2019}
\begin{document}
\maketitle
\begin{abstract}
We examine the possibility that Universe expansion be made of some $\Lambda CDM$ expansions repeating periodically, separated by some inflation and radiation dominated phases. This so-called $\Lambda CDM$ periodic cosmology is motivated by the possibility that inflation and the present phase of accelerated expansion be due to the same dark energy. Then, in a phase space showing the variation of matter density parameter $\Omega_m$ with respect to this of the radiation $\Omega_r$, the curve $\Omega_m(\Omega_r)$ looks like a closed trajectory that Universe could run through forever. In this case, the end of the expansion acceleration of the $\Lambda CDM$ phase is the beginning of a new inflation phase. We show that such a scenario implies the coupling of matter and/or radiation to dark energy. We consider the simplest of these $\Lambda CDM$ periodic models i.e. a vacuum energy coupled to radiation. From matter domination phase to today, it behaves like a $\Lambda CDM$ model, then followed by an inflation phase. But a sudden and fast decay of the dark energy into radiation periodically ends the expansion acceleration. This leads to a radiation dominated Universe preceding a new $\Lambda CDM$ type expansion. The model is constrained with Markov Chain Monte Carlo simulations using supernovae, Hubble expansion, BAO and CMB data and fits the data as well as the $\Lambda CDM$ one.
\end{abstract}
\begin{keywords}
cosmology: theory -- dark energy
\end{keywords}
\section{Introduction}\label{s0}
\begin{figure}
\centering
\includegraphics[width=8cm]{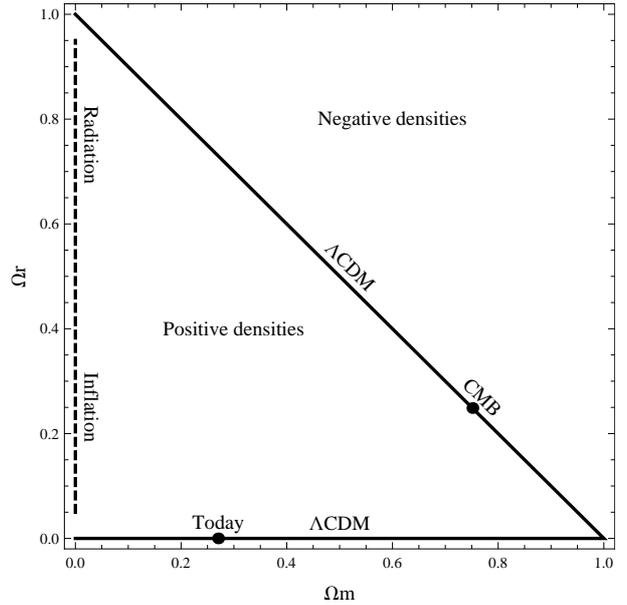}
\caption{\label{fig1}Phase space for the energy density parameters $(\Omega_m,\Omega_r)$ with the constraint $\Omega_m+\Omega_r+\Omega_d=1$. The thick line shows the trajectory of the $\Lambda CDM$ model in agreement with observations and inside the triangular area where the densities are positive. The dashed line is a possible trajectory describing an inflation phase beginning close to the point $\Omega_d\simeq 1$ and followed by a radiation dominated phase going close to the point $\Omega_r\simeq 1$ and different from the $\Lambda CDM$ one. It is then tempting to assume that the whole Universe history could be described by a closed trajectory, i.e. an homoclinic orbit in this phase space.}
\end{figure}

Universe expansion would have accelerated at least two times, a first time during the inflation epoch\citep{Gut81} and a second time during the present epoch\citep{Rie98,Per99}. There are various methods to build a cosmological model with such an expansion history. Hence, in \citet{Sae16}, a modified unimodular gravity version of General Relativity is used that is equivalent to General Relativity at the classical level but also provides a quantum description. In \citet{Li12}, a tachyon condensation on an unstable three-brane gives rise to a tachyon inflation followed by a Chaplygin gas dark matter and dark energy universe. In \citet{PanCha11}, a Chaplygin gas in a spherically symmetric inhomogeneous model is considered for which, at early times, Universe behaves as an Einstein-de Sitter solution generalised to an inhomogeneous spacetime and, at late time, as a $\Lambda CDM$ model. In \citet{PerLim13}, the whole Universe history is described by a vacuum decay.\\
In this paper, a class of cosmological models able to describe the whole Universe expansion history as a succession of $\Lambda CDM$ expansions is found. Let us explain why we consider such a possibility. In a classical way, we assume that Universe is filled by matter, radiation and dark energy with respectively positive densities $\rho_m$, $\rho_r$ and $\rho_d$ (an assumption that could break down near the singularity). Defining their energy density parameters $\Omega_r$, $\Omega_m$ and $\Omega_d$ (see their definitions in the next section) such as $\Omega_m+\Omega_r+\Omega_d=1$, Universe evolution can then be represented as a trajectory in the space phase $(\Omega_m,\Omega_r)$. From today to radiation dominated phases, our Universe is well described by a $\Lambda CDM$ model. Its trajectory in agreement with observations\citep{Hin13} is shown in bold on the phase space of figure \ref{fig1}. One often describes its different epochs by the domination of radiation, matter and then dark energy. But its trajectory also encourages us to describe it as a two phases model, the first one with $\Omega_d\simeq 0$ and the second one with $\Omega_r\simeq 0$. Moreover, at early time, we should have an inflation phase with a very small matter density parameter $\Omega_m\simeq 0$. Inflation is not predicted by the $\Lambda CDM$ model but it could be due to a flat potential (effective, from a scalar field, etc), behaving like a cosmological constant with a value different from the one at our present time (solving the cosmological constant problem\citep{Wei89}). An inflation phase could start with $\Omega_d\simeq 1$ and Universe should evolve to reach the radiation dominated phase $\Omega_r\simeq 1$ (solving the graceful exit problem\citep{Alb82}). This inflation to radiation phase corresponds to the dashed trajectory on figure \ref{fig1}. Considering then this last trajectory associated to the bold $\Lambda CDM$ one, we observe that Universe evolution could thus be described as a close trajectory in the phase space $(\Omega_m,\Omega_r)$, what is named an homoclinic orbit in dynamical system theory. Such a trajectory corresponds to some $\Lambda CDM$ expansions repeating periodically, separated by some inflation and radiation dominated phases (different from the standard $\Lambda CDM$ radiation phase). We call such a periodic behaviour of the expansion a "$\Lambda CDM$ periodic cosmology" and look for the families of cosmological models able to reproduce this scenario of the whole Universe history. Note that we choose to employ the word "periodic" instead of "cyclic" such that there is no confusion with the cyclic cosmological models\citep{SteTur05,BauFra07,Pen06,Ash09} that avoid the beginning of time thanks to some bounces of the scale factor. This is generally not the case of a $\Lambda CDM$ periodic cosmology that has an initial singularity.\\
Note also that we found such a cosmological model in \citet{Fay15} but is was then just a special case for illustration purpose of a subject different from the one of this paper. Here, we want to look for the generic conditions leading to a $\Lambda CDM$ periodic cosmology for various families of cosmological models. This is what is done in the second section of this paper where we examine the case of General Relativity with a dark energy, $F(R)$ or scalar-tensor theories and General Relativity with some viscous or coupled fluids. Among these classes of models, only the last one is able to produce the homoclinic orbits required for a $\Lambda CDM$ periodic cosmology. In a third section, we then study the simplest of these cosmological models consisting in a radiation fluid coupled to a vacuum energy. We constrain it with Markov Chain Monte Carlo simulations applied to supernovae, Hubble expansion, BAO and CMB data. We conclude in the last section by discussing this model from the viewpoints of an effective scalar field potential triggering a warm inflation and an effective fluid unifying dark energy with radiation.
\section{Cosmological models compatible with a $\Lambda CDM$ periodic cosmology}\label{s1}
In this section we look for the cosmological classes of models compatible with the presence of some homoclinic orbits in a phase space $(\Omega_m,\Omega_r)$. This requires the existence of a center equilibrium point in $\Omega_{m}\not =0$ and $\Omega_{r}\not =0$ having pure imaginary eigenvalues\citep{Boy10}. We assume that energy densities of all the species (matter, radiation, dark energy) are positive. 
\subsection{General Relativity and extended theories of gravity}
We first examine General Relativity with non coupled matter, radiation and dark energy. The field equations are
$$
H^2=\frac{k}{3}(\rho_m+\rho_r+\rho)
$$
$$
\frac{d\rho_m}{dt}+3H\rho_m=0
$$
$$
\frac{d\rho_r}{dt}+4H\rho_r=0
$$
$$
\frac{d\rho_d}{dt}+3H(w+1)\rho_d=0
$$
$w=p_d/\rho_d$ is the dark energy equation of state. We define the energy density parameters
\begin{equation}\label{v1}
\Omega_m=\frac{k}{3}\frac{\rho_m}{H^2}
\end{equation}
\begin{equation}\label{v2}
\Omega_r=\frac{k}{3}\frac{\rho_r}{H^2}
\end{equation}
\begin{equation}\label{v3}
\Omega_d=\frac{k}{3}\frac{\rho_d}{H^2}
\end{equation}
The Friedmann equation shows that they are normalised since $\Omega_m+\Omega_r+\Omega_d=1$ and the $\Omega_i$ are positive. We then get the following dynamical system
$$
\Omega_m'=\Omega_m(\Omega_r+3w(1-\Omega_m-\Omega_r))
$$
$$
\Omega_r'=\Omega_r(-1+\Omega_r+3w(1-\Omega_m-\Omega_r))
$$
A prime means a derivative with respect to $N=\ln a$, with $a$ the scale factor. This system is studied in \citet{Fay13} where it is shown that all its equilibrium points are such as $\Omega_m$ or/and $\Omega_r=0$. Hence General Relativity with non coupled matter, radiation and dark energy cannot produce a homoclinic orbit that needs a center equilibrium point with $\Omega_i\not =0$. It is thus not compatible with a $\Lambda CDM$ periodic cosmology. This is also true for any extended theories of gravity that can be rewritten as General Relativity with an effective dark energy\citep{Cap11}. For instance, let us consider an $f(R)$ theory in the Palatini formalism. Using \citet{FayTavTsu07}, we define the following normalised variables
$$
\Omega_m=\frac{k}{3}\frac{\rho_m}{3F\xi H^2}
$$
$$
\Omega_r=\frac{k}{3}\frac{\rho_r}{3F\xi H^2}
$$
$$
\Omega_f=\frac{FR-f}{6F\xi H^2}
$$
with $R$ the usual curvature scalar, $F=df/dR$ and $\xi=\left[1-\frac{3}{2} \frac{dF/dR(FR-2f)}{F(dF/dR R-F)} \right]^2$. Then the field equations rewrite
$$
\Omega_m'=\Omega_m(\Omega_r-(C-3)(1-\Omega_m-\Omega_r))
$$
$$
\Omega_r'=\Omega_r(-1+\Omega_r-(C-3)(1-\Omega_m-\Omega_r))
$$
with $C(R)=-3\frac{(FR-2f)dF/dR R}{(FR-f)(dF/dR R-F)}$. In this phase space, a $\Lambda$CDM periodic cosmology corresponds to the same homoclinic trajectory as described above but with $f\rightarrow 1$ when the expansion mimics a $\Lambda CDM$ model. Identifying $C-3=3w$, we recover the same dynamical system as previously and, once again, we cannot have a center equilibrium point since there is no equilibrium points with $\Omega_i\not =0$. The same could be shown for $f(R)$ theory in metric formalism or scalar-tensor theories, that also reduce to General Relativity with a non coupled dark fluid.
\subsection{Coupled and viscous fluids}
We first consider a dark energy coupled to both radiation and matter with coupling functions $Q_m$ and $Q_r$. We define
$$
q_m=\frac{k}{3}\frac{Q_m}{H^3}
$$
$$
q_r=\frac{k}{3}\frac{Q_r}{H^3}
$$
Then, as shown in \citet{Fay15}, a dynamical system for General Relativity with matter and radiation coupled to dark energy writes 
\begin{equation}\label{eq01}
\Omega_m'=\Omega_m\left[\Omega_r+3w(1-\Omega_m-\Omega_r)\right]+q_m
\end{equation}
\begin{equation}\label{eq02}
\Omega_r'=\Omega_r\left[-1+\Omega_r+3w(1-\Omega_m-\Omega_r)\right]+q_r
\end{equation}
For sake of simplicity, we chose to consider the case $w=-1$ (but our conclusions are the same for any form of $w$). The previous equations system simplifies to
\begin{equation}\label{eq1}
\Omega_m'=\Omega_m(3\Omega_m+4\Omega_r-3)+q_m
\end{equation}
\begin{equation}\label{eq2}
\Omega_r'=\Omega_r(3\Omega_m+4\Omega_r-4)+q_r
\end{equation}
We now look for center equilibrium points $(\Omega_{m(eq)},\Omega_{r(eq)})\not= 0$ responsible for homoclinic orbits. Their eigenvalues that have to be pure imaginary numbers take the form
\begin{eqnarray}
2\lambda_\pm&=&-7+9 \Omega_m+12 \Omega_r+\frac{dq_r}{d\Omega_r}+\frac{dq_m}{d\Omega_m}\pm\surd \mbox{[}1+6 \Omega_m+\nonumber\\
&&9 \Omega_m^2-8 \Omega_r+24 \Omega_m \Omega_r+16 \Omega_r^2+\frac{dq_r}{d\Omega_r}^2+2 \frac{dq_m}{d\Omega_m}+\nonumber\\
&&6 \Omega_m \frac{dq_m}{d\Omega_m}-8 \Omega_r \frac{dq_m}{d\Omega_m}+\frac{dq_m}{d\Omega_m}^2-2 \frac{dq_r}{d\Omega_r} (1+3 \Omega_m-\nonumber\\
&&4 \Omega_r+\frac{dq_m}{d\Omega_m})+16 \Omega_m \frac{dq_r}{d\Omega_m}+4 \frac{dq_m}{d\Omega_r} (3 \Omega_r+\frac{dq_r}{d\Omega_m})\mbox{]}
\end{eqnarray}
We can show with a numerical example that, depending on the forms of $q_m$ and $q_r$, center equilibrium points are possible for this phase space: 
\begin{itemize}
\item Choose some numerical values $\Omega_{m(eq)}>0$ and $\Omega_{r(eq)}>0$ with $\Omega_{m(eq)}+\Omega_{r(eq)}<1$. 
\item Then calculate $q_m$ and $q_r$ such as $\Omega_m'$ and $\Omega_r'=0$ in $(\Omega_{m(eq)},\Omega_{r(eq)})$
\item Choose a numerical value for $\frac{dq_m}{d\Omega_m}(\Omega_{m(eq)},\Omega_{r(eq)})$
\item Then calculate $\frac{dq_r}{d\Omega_r}$ such as $-7+9 \Omega_m+12 \Omega_r+\frac{dq_r}{d\Omega_r}+\frac{dq_m}{d\Omega_m}=0$ in $\Omega_{m(eq)}$ and $\Omega_{r(eq)}$.
\item Choose some numerical values for $\frac{dq_m}{d\Omega_r}(\Omega_{m(eq)},\Omega_{r(eq)})$ and $\frac{dq_r}{d\Omega_m}(\Omega_{m(eq)},\Omega_{r(eq)})$ such as the above square root is negative in $(\Omega_{m(eq)},\Omega_{r(eq)})$.
\end{itemize}
This shows that it is possible to choose appropriate values for $\Omega_{m(eq)}$, $\Omega_{r(eq)}$, $\frac{dq_m}{d\Omega_m}(\Omega_{m(eq)},\Omega_{r(eq)})$, $\frac{dq_m}{d\Omega_r}(\Omega_{m(eq)},\Omega_{r(eq)})$ and $\frac{dq_r}{d\Omega_m}(\Omega_{m(eq)},\Omega_{r(eq)})$ such as some coupling functions allow homoclinic orbits around the center point $(\Omega_{m(eq)},\Omega_{r(eq)})$. Of course, when one assumes the forms of $q_m$ and $q_r$, this adds some constraints that render possible or not the above process. In the next subsections, we consider the special cases $q_m=0$ and $q_r=0$ for any form of $w$.
\subsubsection{$q_m=0$}\label{s11}
Here, we consider any forms for $w$. There is only one possible center equilibrium point that writes
$$
(\Omega_{m(eq)},\Omega_{r(eq)})=(\frac{q_r+3 w-3 q_r w}{3 w},q_r)
$$
Then $(\Omega_{m(eq)}>0,\Omega_{r(eq)}>0)$ and $\Omega_{m(eq)}+\Omega_{r(eq)}<1$ if
$$
w<0
$$
and
$$
0<q_r<\frac{3w}{3w-1}
$$
The conditions such as $\lambda_\pm$ be pure imaginary numbers are that
$$
\frac{dq_r}{d\Omega_r}=\frac{3 w^2 (1+3 w)+3 q_r^2 w \frac{dw}{d\Omega_r}+q_r (q_r+3 w-3 q_r w) \frac{dw}{d\Omega_m}}{3 w^2}
$$
and $\frac{dq_r}{d\Omega_m}>$ ($<$) to
$$
\frac{\left(3 w^2+q_r \frac{dw}{d\Omega_m}\right) \left(9 w^3+3 q_r^2 w \frac{dw}{d\Omega_r}+q_r (q_r+3 w-3 q_r w) \frac{dw}{d\Omega_m}\right)}{3 w^2 \left(w (-1+3 w)+q_r \frac{dw}{d\Omega_r}\right)}
$$ 
when $q_r \frac{dw}{d\Omega_r}>w(1-3w)$ (respectively $<w(1-3w)$) in $(\Omega_{m(eq)},\Omega_{r(eq)})$. Note that if we set $w=-1$, there should thus be some homoclinic trajectories corresponding to a $\Lambda CDM$ periodic model. However in \citet{Fay14}, we show that when radiation is coupled to vacuum energy, an inflation phase needs a negative radiation density at early time whereas here, we assume $\rho_r>0$. This apparent contradiction comes from the fact that in \citet{Fay14}, we assumed a source point to represent the beginning of the inflation phase. But this is no more required in the present paper where we consider homoclinic orbits and we can thus have an inflation phase with $q_m=0$ in the case of a periodic $\Lambda CDM$ model with positive densities.
\subsubsection{$q_r=0$}\label{s12}
Here again, we do not specify $w$. Then the only possible center equilibrium point is located in 
$$
(\Omega_{m(eq)},\Omega_{r(eq)})=(-q_m,\frac{-1+3 w+3 q_m w}{-1+3 w})
$$
Then $(\Omega_{m(eq)}>0,\Omega_{r(eq)}>0)$ and $\Omega_{m(eq)}+\Omega_{r(eq)}<1$ if
$$
w>1/3
$$
that discards the special constant case $w=-1$\citep{Fay14}, and 
$$
q_1>\frac{1}{3w}-1
$$
$q_1$ is thus negative, implying an energy transfer from dark matter to dark energy. The eigenvalues $\lambda_\pm$ are pure imaginary numbers if
\begin{eqnarray}
\frac{dq_m}{d\Omega_m}=&&(3 q_m (-1+3 (1+q_m) w) \frac{dw}{d\Omega_r}+(3 w-1)\times\nonumber\\
&& (2-9 w+9 w^2-3 q_m^2 \frac{dw}{d\Omega_m}))(1-3 w)^{-2}\nonumber\\\nonumber
\end{eqnarray}
and $\frac{dq_m}{d\Omega_r}<$ or $>$
\begin{eqnarray}
((1-3 w)^2+3 q_m \frac{dw}{d\Omega_r}) (3 q_m (-1+3 (1+q_m) w) \frac{dw}{d\Omega_r}+\nonumber\\
(-1+3 w) ((1-3 w)^2-3 q_m^2 \frac{dw}{d\Omega_m}))\nonumber\\
(3 (1-3 w)^2 (w (-1+3 w)+q_m \frac{dw}{d\Omega_m}))^{-1}\nonumber\\\nonumber
\end{eqnarray}
when $q_m \frac{dw}{d\Omega_m}>w(1-3w)$ (respectively $<w(1-3w)$) in $(\Omega_{m(eq)},\Omega_{r(eq)})$.
\subsubsection{Viscous fluid}\label{s13}
A viscous fluid can be defined by adding to the usual definition of pressure a term $B_i$ containing a viscosity coefficient(see for instance \citet{BaoBar09}). For instance, if dark energy is a viscous fluid, we then have $p_d=w_d\rho_d-B_d$. In the extreme case where all the fluids are viscous, then defining $q=k(B_m+B_r+B_d)/H^2$, the dynamical system for viscous fluid is found by putting in equations (\ref{eq01}-\ref{eq02}) $q_m=-q \Omega_m$ and $q_r=-q \Omega_r$. The equilibrium points are 
$$
((0,-\frac{1+q-3 w}{-1+3 w}),(-\frac{q-3 w}{3 w},0),(0,0))
$$
Since all of them have a vanishing $\Omega_m$ or $\Omega_r$, there is thus no homoclinic orbit for General Relativity with viscous fluids.\\
\\
Hence, if we assume positive energy densities, a $\Lambda CDM$ periodic model can only be reproduced by a dark energy coupled to radiation or/and matter. In the next section, we examine the simplest $\Lambda CDM$ periodic model and constrain it with some observations.
\section{The simplest $\Lambda CDM$ periodic model}\label{s2}
The simplest $\Lambda CDM$ periodic model is defined by
$$
w=-1
$$
$$
q_m=0
$$
$$
q_r=\alpha \Omega_r\Omega_d
$$
with $\alpha$, a positive constant. A dark energy coupled to radiation is not new. It is used in warm inflation paradigm\citep{Ber98} for instance. In what follows, we examine the dynamics of this model (expansion, densities, coupling function, etc) and how observations constrain it.
\subsection{Model dynamics}\label{s21}
\begin{figure}
\centering
\includegraphics[width=8cm]{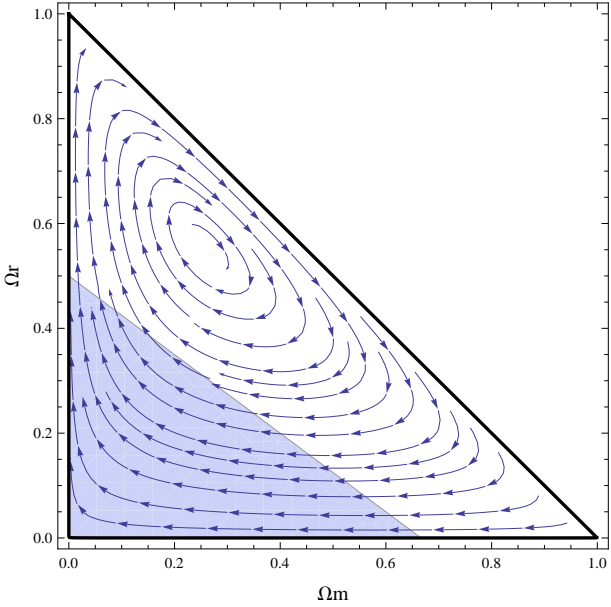}
\caption{\label{fig2}Phase space for a $\Lambda$CDM periodic model defined by $w=-1$, $q_m=0$ and $q_r=\alpha \Omega_r\Omega_d$ with $\alpha=5.42$. The gray part is the set of points for which Universe expansion is accelerated.}
\end{figure}

\begin{figure}
\centering
\includegraphics[width=8cm]{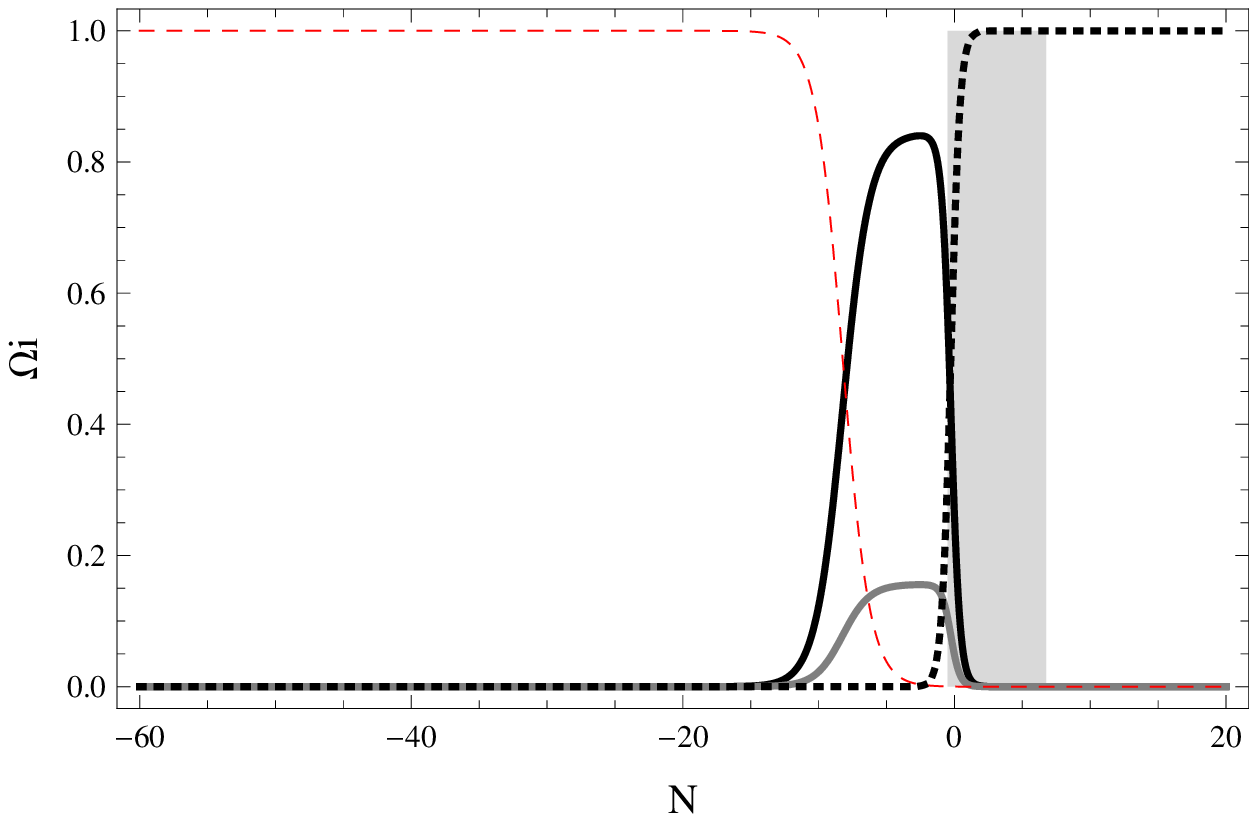}
\includegraphics[width=8cm]{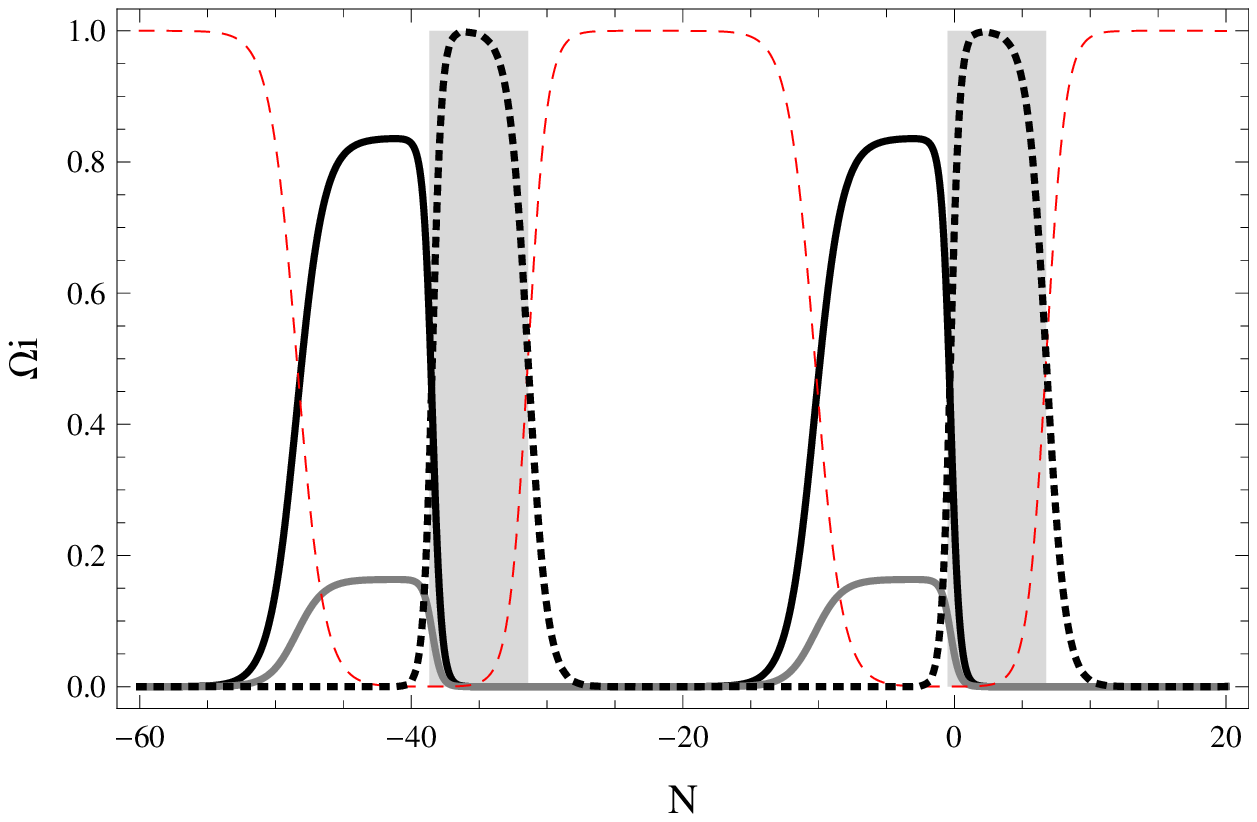}
\caption{\label{fig4}Density parameters $\Omega_{CDM}$ (thick, black), $\Omega_{b}$ (thick, gray), $\Omega_r$ (dashed) and $\Omega_d$ (dotted) for the $\Lambda CDM$ (first figure) and coupled models (second figure, with $\alpha=5.42$, see subsection \ref{s22}). The gray rectangles indicate an acceleration of the expansion. The density parameters evolve in the same way from today ($N=0$) to the matter domination phase ($N\simeq -1.8$, i.e $z\simeq 5$). Then, in agreement with the phase space on figure \ref{fig2}, the density parameters of the coupled model have a periodic behaviour with a period $\Delta N = 38.1$.}
\end{figure}

\begin{figure}
\centering
\includegraphics[width=8cm]{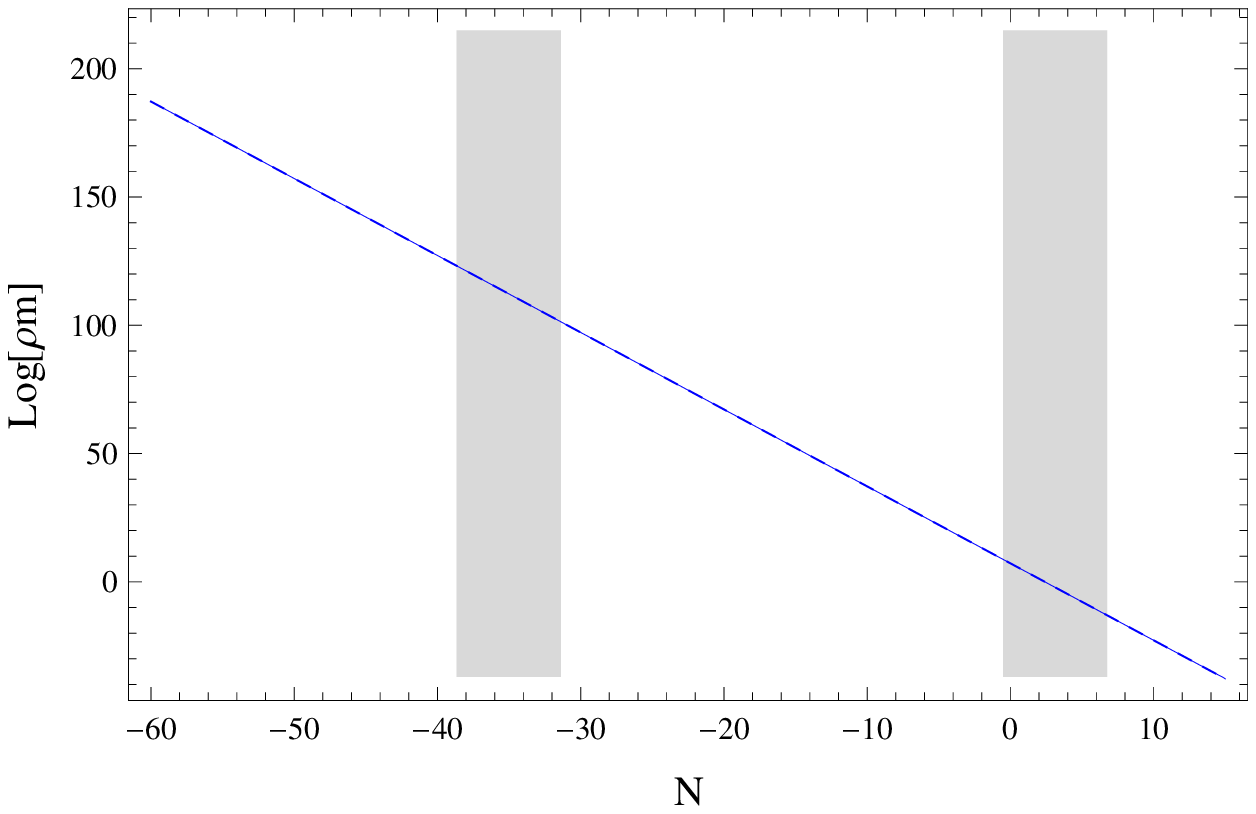}
\includegraphics[width=8cm]{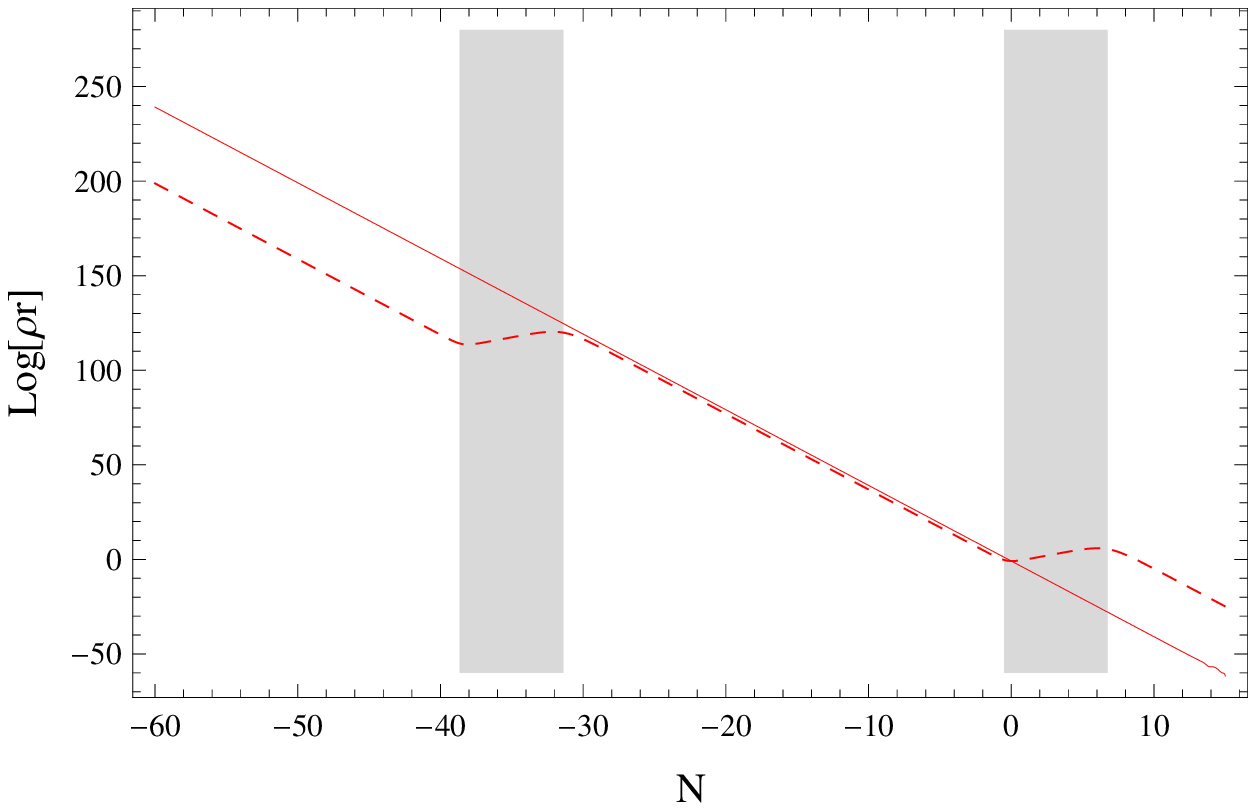}
\includegraphics[width=8cm]{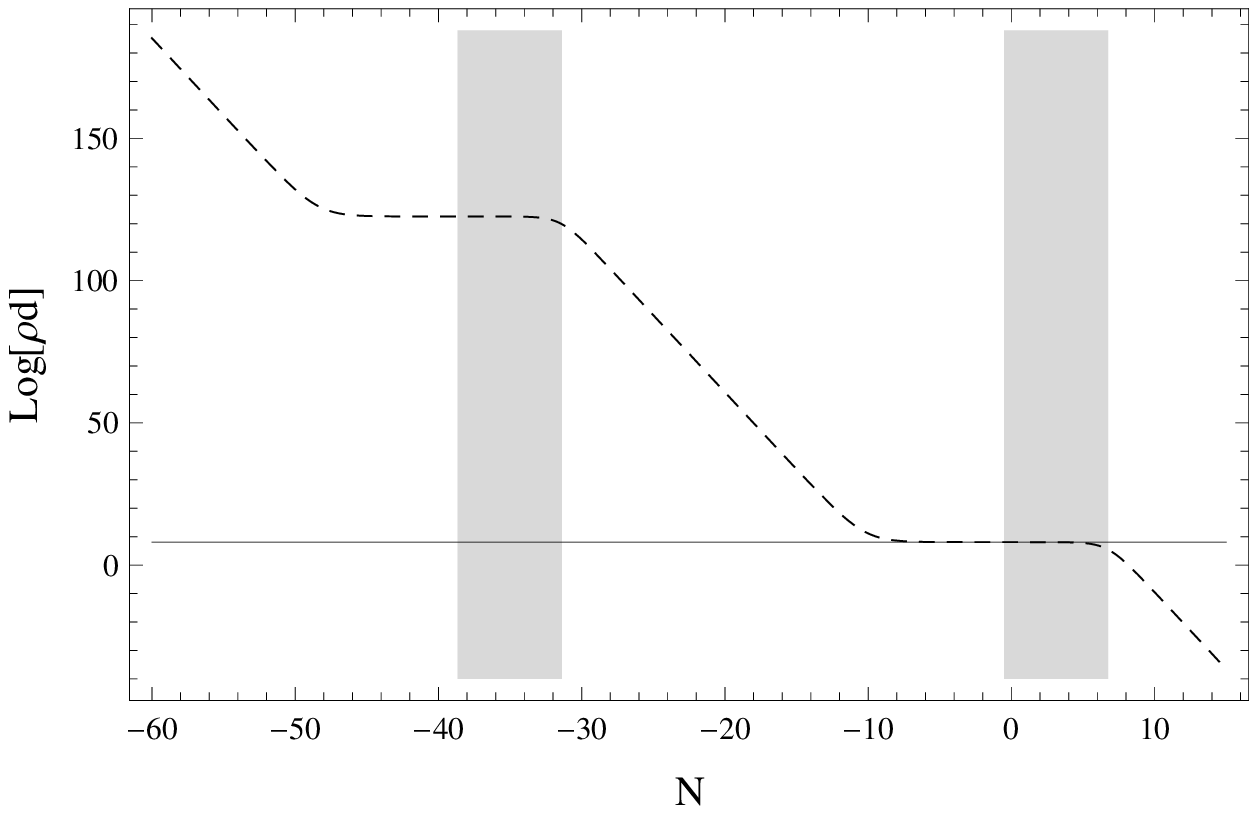}
\includegraphics[width=8cm]{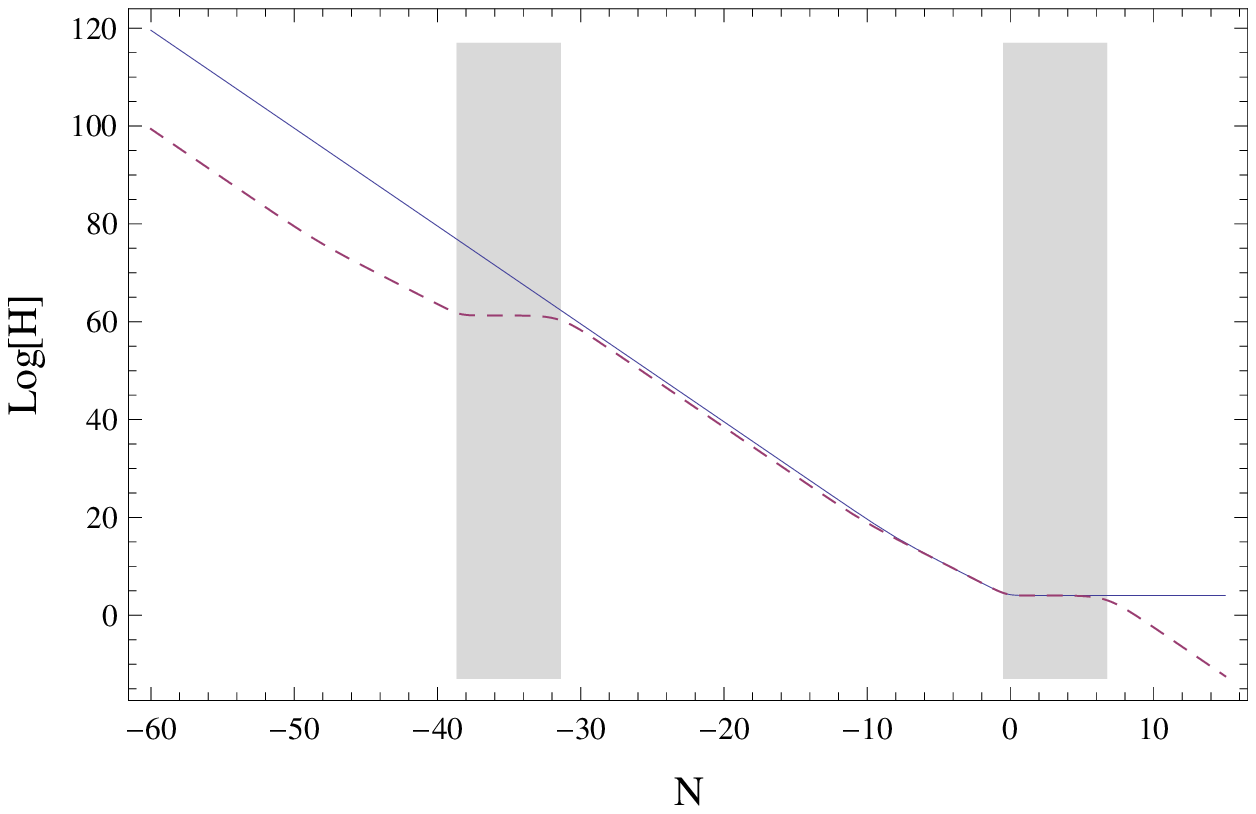}
\caption{\label{fig5}First, second and third graphs show the matter, radiation and dark energy densities for the $\Lambda CDM$ (thin) and coupled($\alpha=5.42$, dashed) models. The fourth graph shows the Hubble function. Gray area are phases of accelerated behaviour for the coupled model.}
\end{figure}

\begin{figure}
\centering
\includegraphics[width=8cm]{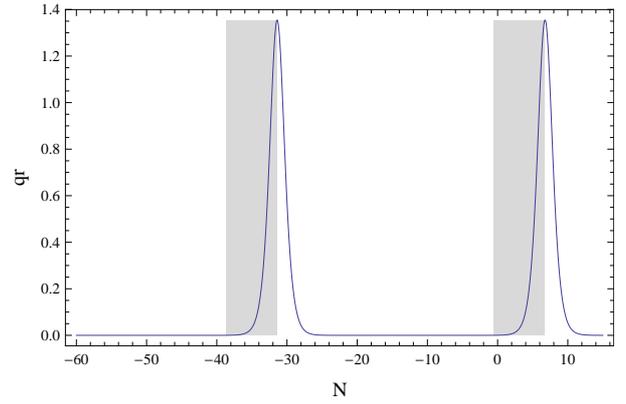}
\caption{\label{fig6}$q_r(N)$ and some periods of accelerated expansion in gray.}
\end{figure}

The equilibrium points with non vanishing coordinates write
$$
(\Omega_m,\Omega_r)=(\frac{1}{3}(3-4\alpha(1-\Omega_m-\Omega_r)\Omega_r),\alpha(1-\Omega_m-\Omega_r)\Omega_r)
$$
The eigenvalues take the form
$$
\lambda_\pm=\lambda_1\pm\frac{\sqrt{3}}{6}\sqrt{\lambda_2}
$$
with 
$$
\lambda_1=1-\frac{1}{2}\alpha(\Omega_m+2\Omega_r-1)
$$
\begin{eqnarray}
\lambda_2&=&48+24 \alpha  \left(-1+\Omega_m-2 \Omega_r+2 \Omega_m \Omega_r+2 \Omega_r^2\right)+\nonumber\\
&&\alpha ^2 (3+36 \Omega_r-68 \Omega_r^2+32 \Omega_r^3+\Omega_m^2 (3+48 \Omega_r)+\nonumber\\
&&\Omega_m (-6-84 \Omega_r+80\Omega_r^2))\nonumber
\end{eqnarray}
They are pure imaginary numbers if $\lambda_1=0$. This implies
$$
\alpha=\frac{2}{\Omega_m+2\Omega_r-1}
$$
Introducing this value for $\alpha$ in the above equilibrium points, we derive that the set of $(\Omega_m,\Omega_r)$ equilibrium points in agreement with an homoclinic orbit is on the line
$$
\Omega_r=\frac{3}{4}(1-\Omega_m)
$$
and thus $\alpha>4$. Introducing this expression for $\Omega_r$ in the eigenvalues, it comes 
$$
\lambda_\pm=\pm\sqrt{-3\Omega_m}
$$
which are thus pure imaginary. A phase space with $\alpha=5.42$ (in agreement with the data, see subsection \ref{s22}) is plotted on figure \ref{fig2}. The density parameters $\Omega_i$ for the $\Lambda CDM$ and coupled models are shown on figure \ref{fig4}. Note that we split $\Omega_m$ in its two components for cold dark matter and baryon, i.e. $\Omega_m=\Omega_{CDM}+\Omega_b$ since in subsection \ref{s22} we use some BAO and CMB observations to constrain the model and the sound horizon depends on the baryons density.
The density parameters $\Omega_i$ of the two models evolve very closely from today until $N\simeq -1.8$ or $z\simeq 5$, i.e until the matter domination phase. For smaller values of $N$, both models behave differently. In particular, in agreement with phase space of figure \ref{fig2}, figure \ref{fig4} shows the periodic behaviour of the coupled model. It has a singularity in $N\rightarrow -\infty$ and expands to the infinite future in $N\rightarrow +\infty$.\\
Let us examine the behaviours of densities, Hubble function and coupling function $q_r$. They are shown on figures \ref{fig5} and \ref{fig6}. To clarify the discussion, we define a period as the interval of time during which a trajectory leaves and comes back near the phase space point $(\Omega_m,\Omega_r)=(0,0)$ or $\Omega_d\simeq 1$. For $\alpha=5.42$, a period lasts $\Delta N=38.1$.
\begin{itemize}
\item Matter is not coupled. Its density $\rho_m$ thus behaves the same way in the coupled and $\Lambda CDM$ models.
\item Dark energy is a decreasing function. Its density $\rho_d$ periodically becomes nearly a constant, mimicking a cosmological constant. Then, when $\rho_d$ is nearly a constant whereas matter and radiation densities decreases, dark energy ends up dominating Universe content, accelerating its expansion. The past phases of acceleration can thus be considered as an infinite number of inflation phases. We are presently at the beginning of a new acceleration phase during which radiation density will increase as shown by the peak in the coupling function on figure \ref{fig6}: dark energy will be cast into radiation that will dominate Universe content in the next period, ending the present acceleration phase.
\item Radiation decreases but during the acceleration of the expansion. Then as shown by the increase of $q_r$ on figure \ref{fig6}, dark energy is cast into radiation and Universe becomes progressively radiation dominated with a decelerated expansion. This does not mean that the next period will experiment a hot phase as the previous one when Universe was very small, just that radiation will dominate a cold Universe.
\item The Hubble function approximates a $\Lambda CDM$ expansion when matter dominates ($\Omega_m\simeq 1$) to the end of the period (i.e. $(\Omega_m,\Omega_r)\simeq (0,0)$). At each period, the value of $\Lambda$ is smaller and smaller. The value of $\Lambda$ necessary to explain the present Universe acceleration is thus different from the value necessary to explain inflation, that is the cosmological constant problem.
\item As shown on figure \ref{fig6}, the coupling function $q_r$ starts to increase quickly when dark energy dominates Universe and accelerate its expansion. Then dark energy is cast into radiation whose density increases and comes to dominate Universe. This ends the expansion acceleration and anew, $\rho_d$ and $\rho_r$ decrease as well as $q_r$ that quickly becomes small, allowing the Hubble function to progressively converge to a $\Lambda CDM$ behaviour after dark energy density reached a new constant value.
\end{itemize}
\subsection{Observational constraints}\label{s22}
\begin{figure*}
\centering
\includegraphics[width=14cm]{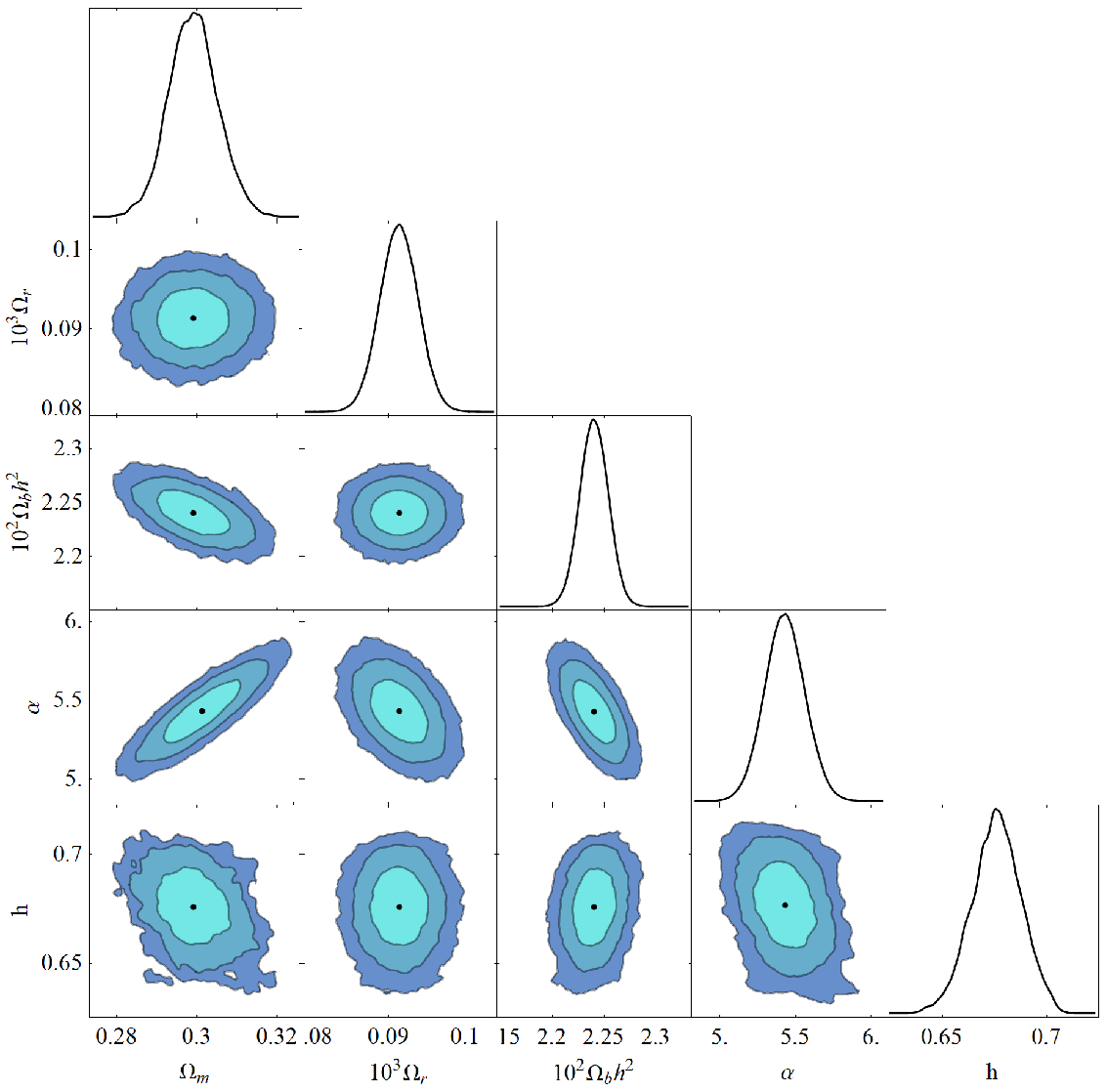}
\caption{\label{fig13}$68.3\%$, $95.4\%$ and $99.7\%$ confidence contours and one-dimensional marginalized likelihoods for the coupled model $q_m=0$ and $q_r=\alpha \Omega_r\Omega_d$. The black points are the mean MCMC values $(\Omega_{m0},10^3\Omega_{r0},\Omega_{b0}h_0^2,\alpha,H_0)=(0.291,0.091,0.0224,5.43,67.58)$.}
\end{figure*}

We now check if this model is in agreement with some observational data. We use a Markov Chain Monte Carlo (MCMC) sampler\citep{Arj19A, Arj19} whose code is available at members.ift.uam-csic.es/savvas.nesseris/. We consider the following data (where a subscript $0$ means a value of a parameter today)\\
\begin{itemize}
\item For the CMB, we consider the shift parameters $(R, l_a)$ and $\Omega_{b0}h_0^2$, with $H_0=100h_0$, based on Planck final release\citep{Agh18} as derived by \citet{Zha19}.
\item For the CMB, we also consider its temperature that constrains $\Omega_{r0}$ and avoids its degeneracy with the $\alpha$ parameter. $\Omega_{r0}$ can be separated in two components, one for the photons $\Omega_{\gamma 0}$ and one for the relativistic neutrinos $\Omega_{\nu 0}$. The photons density $\Omega_{\gamma 0}h_0^2$ is precisely measured by the CMB temperature as $\Omega_{\gamma 0}h_0^2=2.47282\times 10^{-5}$. $\Omega_{\nu 0}$ is related to $\Omega_{\gamma 0}$ by(see \cite{Kom09} for instance) $\Omega_{\nu 0}=0.2271\Omega_{\gamma 0} N_{eff}$ where $N_{eff}$ is the effective number of neutrino species. Then, the values $h_0=0.674\pm0.005$ and $N_{eff}=2.99\pm 0.17$ by Planck\citep{Agh18} lead to the $1\sigma$ constraint $\Omega_{r0}=(9.13\pm 0.25)\times 10^{-5}$ that we introduce as a prior in the $\chi^2$ defined below.
\item For the supernovae, we consider the $1048$ supernovae from the last Pantheon data\citep{Sco18}.
\item For the BAO measurements, we consider the Ly$\alpha$ forest from BOSS DR11\citep{Del15}, BOSS DR12\citep{Gil16}, WiggleZ\citep{Bla12}, 6dFGS\citep{Beu11} and DES Year 1\citep{Abb18}. This corresponds to $10$ measurements.
\item For the Hubble function, we consider $36$ data points from \citet{Mor12A,Mor12B,Mor15,Mor16} and \cite{Guo16}.
\end{itemize}
The total $\chi^2$ is defined as $\chi^2=\chi_{CMB}^2+\chi_{SN}^2+\chi_{BAO}^2+\chi_{H}^2+(\Omega_{r0}-9.13\times 10^{-5})^2/(0.25\times 10^{-5})^2$. The priors used for the MCMC analysis are $\Omega_{m0}\in\left[0.2,0.4\right]$, $\Omega_{r0}\in\left[10^{-5},10^{-4}\right]$, $\Omega_{b0}h_0\in\left[0.001,0.06\right]$ and $h_{0}\in\left[0.55,0.80\right]$. Figure \ref{fig13} shows the $68.3\%$, $95.4\%$ and $99.7\%$ confidence contours for the coupled model and the one-dimensional marginalized likelihoods got with a sample of $500 000$ MCMC points (the black points are the mean MCMC values).At $68.3\%$, we have $\Omega_{m0}=0.298\pm 0.006$, $10^3\Omega_{r0}=0.091\pm 0.002$, $\Omega_{b0}h_0^2=0.0224\pm 0.0001$, $\alpha=5.42\pm 0.13$ and $H_0=67.74\pm 1.17$, the best $\chi^2$ being $1068.07$. The best $\Lambda CDM$ and coupled models $\chi^2$ with their components are shown in table \ref{tab1}. Using the Akaike information criterion (AIC)\citep{Aka74} as done in \citet{Arj19A} and \citet{Arj19}, we can compare the two models. Both of them being constrained by the same $1098$ data and the coupled model having one more free parameter than the $\Lambda CDM$ one, we find that 
$$
AIC_{coupled}-AIC_{\Lambda CDM}\simeq 2.7
$$
Interpreted with the Jeffreys' scale, this last value should mean that both models are statistically equivalent. This last conclusion has to be considered with care as it is shown in \citet{Nes13}.\\
\begin{table}
\begin{center}
\begin{tabular}{|c||c|c|c|c|c|}
\hline
 & $\chi_{CMB}^2$ & $\chi_{SN}^2$ & $\chi_{BAO}^2$ & $\chi_{H}^2$ & $\chi^2$ \\
 \hline
$\Lambda CDM$ & $0.59$ & $1036.42$ & $13.50$ & $22.34$ & $1072.86$ \\
Coupled model  & $0.43$ & $1035.97$ & $10.56$ & $21.10$ & $1068.07$\\
\hline
\end{tabular}
\caption{$\chi^2$ for the $\Lambda CDM$ and best fit coupled model with $\alpha=5.42$}
\label{tab1}
\end{center}
\end{table}
Let us take $\alpha=5.42$, the best fitting value for $\alpha$. Then, a period lasts $\Delta N = 38.1$. The accelerated expansion phases repeat each time dark energy is dominating and each of them lasts $N=7.3$. Universe age is $13.9$ billions years old, as for the $\Lambda CDM$ model. This is also approximately the duration of the present period from its beginning around $N\simeq -36.0$ to today. It will end in $N\simeq 2.1$, i.e. in $35$ billions years. The previous period only lasts $5$ years. The periods are thus (with respect to proper time $t$) shorter and shorter as we go into the past. This also means that during the last period, no structure had time to form and thus cannot be seen in the CMB. 
\section{Discussion and conclusion}\label{s3}
\begin{figure}
\centering
\includegraphics[width=8cm]{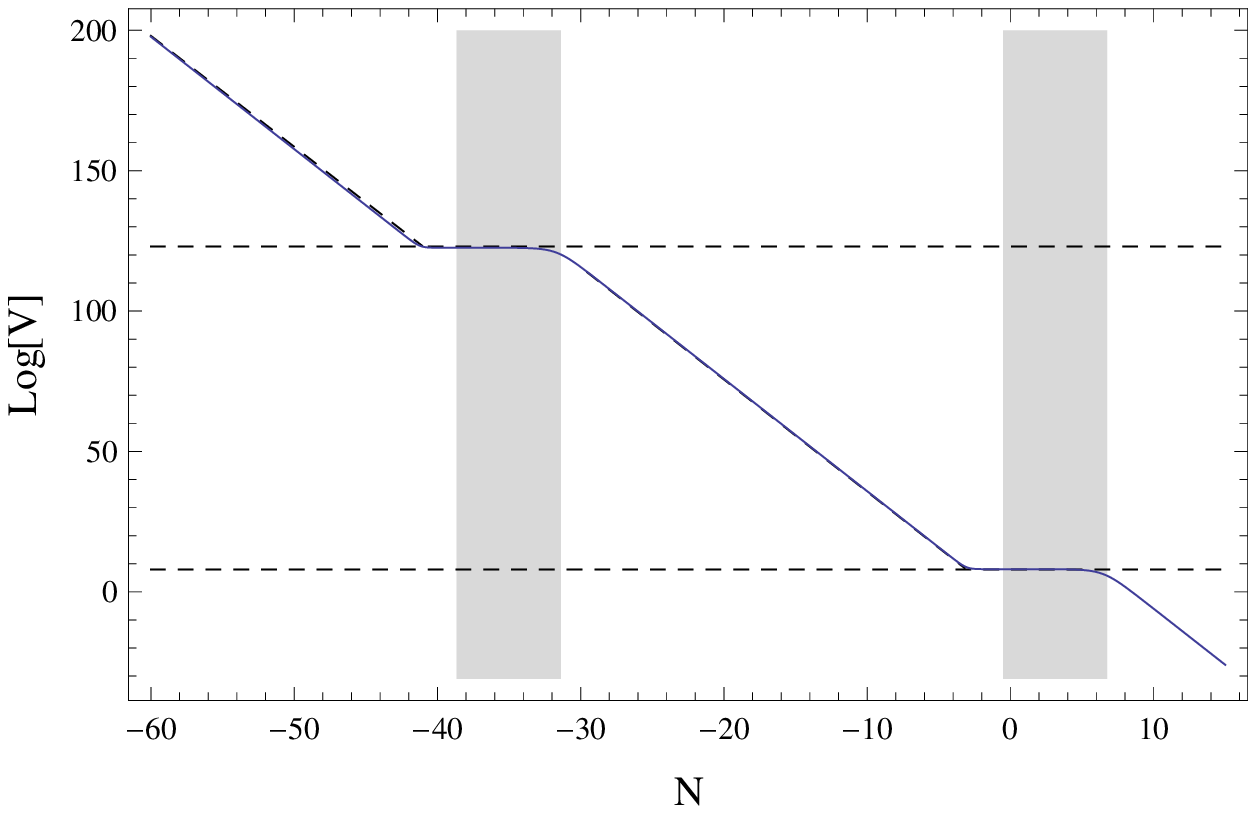}
\includegraphics[width=8cm]{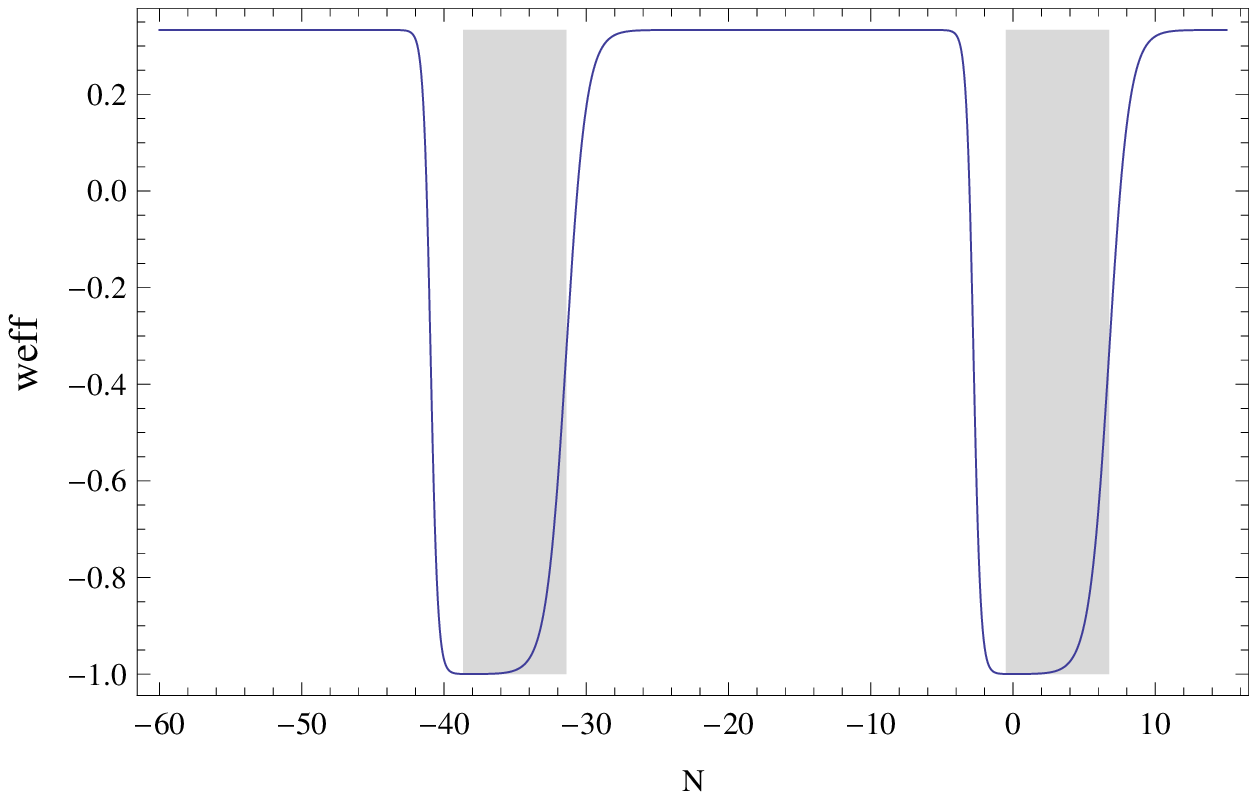}
\caption{\label{fig7}Effective potential (the dashed lines approximate the potential) and equation of state of the effective fluid $\rho_{eff}=\rho_d+\rho_r$.}
\end{figure}

This paper starts from the idea that Universe expansion is well described by a $\Lambda CDM$ model, preceded by an inflation phase at early time that could also be due to a vacuum energy or a very flat potential when matter and radiation were negligible. In the phase space $(\Omega_m,\Omega_r)$, such a description looks like the bold and dashed trajectories on figure \ref{fig1}. This suggests that such a scenario could be described by an homoclinic orbit. The behaviours of the density parameters $\Omega_i$ would thus repeat periodically as well as a $\Lambda CDM$ type expansion with different values of $\Lambda$ (see below), hence the name $\Lambda CDM$ periodic cosmology.\\
Among the classes of models we consider in section \ref{s1}, the only one in agreement with homoclinic orbits is General Relativity with a dark energy coupled to matter and/or radiation. We then defined the conditions on the dark energy equation of state $w$ and the coupling functions $q_m$ and $q_r$ to get some homoclinic orbits. Then, we considered the simplest of these models defined by $w=-1$, $q_m=0$ and $q_r=\alpha \Omega_r\Omega_d$. The conditions for homoclinic orbits impose $\alpha>4$, meaning that the model cannot be reduced to a $\Lambda CDM$ model with $\alpha=0$. This corresponds to a vacuum energy coupled to the radiation, the former decaying into the latter.\\
We constrained this model with Markov Chain Monte Carlo simulations using supernovae, Hubble expansion, BAO and CMB data. Then, we found at $68.3\%$ that $\Omega_{m0}=0.298\pm 0.006$, $10^3\Omega_{r0}=0.091\pm 0.002$, $\Omega_{b0}h_0^2=0.0224\pm 0.0001$, $\alpha=5.42\pm 0.13$ and $H_0=67.74\pm 1.17$.\\
Choosing the best fit value $\alpha=5.42$, a period of this constrained cosmological model, defined as the interval of time during which a trajectory leaves and comes back near the phase space point $(\Omega_m,\Omega_r)=(0,0)$, lasts $\Delta N=38.1$ and can be described as follows. It begins when density parameters of the matter and radiation are very small. Dark energy behaves then like a cosmological constant. The coupling between dark energy and radiation is stronger and stronger. More and more dark energy is cast into radiation that comes to dominate. The expansion acceleration ends whereas Universe becomes dominated by radiation. The densities of all the species then decrease. The density of dark energy reaches a new constant value and the expansion dominated by matter begins to behave as a $\Lambda CDM$ expansion. Consequently, dark energy dominates again and a new phase of accelerated expansion takes place whereas the radiation density starts to increase, beginning a new (colder) period. Note that our period is the first one long enough for life and structures to appear. It begins $13.9$ billions years ago (nearly the Universe age) and will end in $35$ billions years.\\\\
We conclude this paper by examining the periodic $\Lambda CDM$ model from the viewpoints of an effective scalar field potential triggering a warm inflation and an effective fluid unifying dark energy with radiation.\\
From the viewpoint of inflation, periodic $\Lambda CDM$ models are examples of warm inflation\citep{Ber95} since radiation is produced during the inflation phase by the decay of dark energy. Hence the inflationary phase ends into a radiation dominated phase, avoiding the graceful exit problem\citep{GutWei83}. For the specific model of section \ref{s2}, it is interesting to look for an effective scalar field $\phi$ with a density $\rho_{eff}=\rho_d+\rho_r$. Its effective potential $V_{eff}$ is plotted on the first graph of figure \ref{fig7} as a function of $N$. When expansion is accelerated, the scalar field potential $V_{eff}$ behaves as a cosmological constant $\Lambda_{eff}$ with smaller and smaller value at each period as we go to the future. Comparing the scalar field potential on figure \ref{fig7} with the radiation density parameter on figure \ref{fig4}, we note that each time a slow roll\citep{Lin82, Alb82} ends, a reheating starts. With $\alpha = 5.42$, we have from one period $i$ to the next one $i+1$, $\Lambda_{eff(i+1)}\simeq \Lambda_{eff(i)})e^{-114.66}$. When $V_{eff}$ is not a constant, $\ln(V_{eff})$ can be approximated by a straight line with a slope $-4$. Note that one gets a similar figure for $V_{eff}(\phi)$, but with slightly curved instead of straight lines.\\
The meaning of this slope is clarified when considering unification of dark energy and radiation from the viewpoint of an effective fluid with equation of state $w_{eff}=p_{eff}/\rho_{eff}$. It is plotted on the second graph on figure \ref{fig7}. Since $w_{eff}\geq -1$, the problem of a ghost-like dark energy\citep{Cal03} is avoided. The effective fluid alternatively behaves like a vacuum energy ($w=-1$ and the scalar field potential $V_{eff}$ is constant) or a radiation fluid ($w=1/3$ and the potential decreases as $e^{-4N}$, i.e. $a^{-4}$ like during a standard radiation dominated phase). The transition between these two values for $w_{eff}$ corresponds to the rise of the radiation density parameter (transition from $w_{eff}=-1$ to $1/3$) and the rise of the dark energy density parameter (transition from $w_{eff}=1/3$ to $-1$).\\\\
The specific $\Lambda CDM$ periodic model we studied in this paper is the simplest of this class of models. It does not pretend to solve all the cosmological problems. Hence, one could wish longer phases of inflation to get more e-fold at each period (for instance by choosing different forms of the coupling functions) or avoid the Big Bang singularity (which possibly could be done by reversing the sign of $H$ on an homoclinic orbit). It would also be interesting to look for a specific signature of such a periodic expansion, maybe in the CMB.

\section*{Acknowledgement}
I thank Professor Savvas Nesseris for his Mathematica codes for the Markov Chain Monte Carlo simulations and the referee who suggests to do such an analysis.

\bibliographystyle{unsrt}

\end{document}